\documentclass[twocolumn,prl,aps,showpacs]{revtex4}

\usepackage{graphicx}
\usepackage{dcolumn}

\begin{document}

\title{Proposal of Two-Pulse Excitation to Efficiently Photoinduced\\
Novel $sp^3$ Nano Domain with Frozen Shear in Graphite Crystal}
\author{Keita Nishioka$^1$}
\author{Keiichiro Nasu$^2$}
\author{Kenji Yonemitsu$^{1,3}$}
\address{$^1$Institute for Molecular Science, Okazaki, Aichi 444-8585, Japan\\
$^2$Solid State Theory Division, Institute of Materials Structure Science, High Energy Accelerator Research Organization (KEK), Tsukuba, Ibaraki, 305-0801, Japan\\
$^3$Department of Functional Molecular Science, Graduate University for Advanced Studies, Okazaki, Aichi 444-8585, Japan}
\date{\today}

\begin{abstract}
We propose a two-pulse excitation to efficiently photoinduced a novel $sp^3$-bonded nano-domain with a frozen shear in a graphite crystal. This $sp^3$ structure is known to be well stabilized by shear displacement between neighboring two graphite layers. This shear motion is generated only as a transient and unfrozen one by the first visible laser pulse, shone over the graphite crystal. While, the second pulse is proved to freeze it before it disappears, resulting in an efficient interlayer $\sigma$ ($sp^3$) bond formation.
\end{abstract}

\pacs{64.70.Nd,61.48.Gh,31.15.xv,05.40.-a}

\maketitle

Plenty of experimental and theoretical studies for photoinduced structural phase transitions (PSPT) have revealed the properties of multi-stability hidden in various materials\cite{Nasu1,Nasu2,Yonemitsu,Koshihara,Nasu3}. Carbon is one of such materials and exhibits various condensed phases with $sp^2$ and $sp^3$ structures\cite{Kroto,Iijima,Tateyama}. As well known, graphite, which is the most stable phase of condensed carbon systems, has a layer structure characterized by an $sp^2$-bonded network.

\begin{figure}
\begin{center}
  \includegraphics[width=0.8\linewidth]{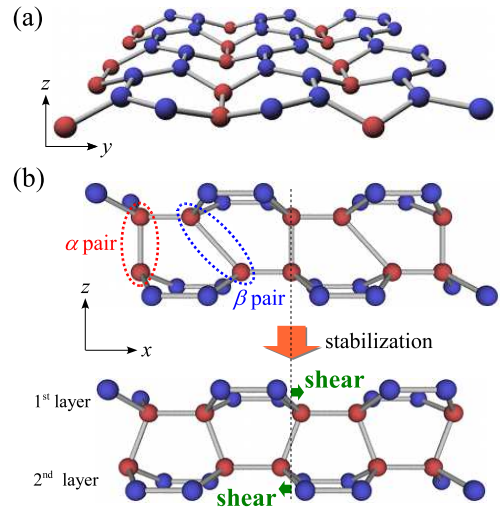}
  \caption{The structure of diaphite. (a) The top-down view of the first layer. The blue spheres indicate rising up carbons from the layer and the red ones are sinking down carbons. (b) The side view of the diaphite. When the first and the second layers just approach each other, the interlayer bonds in $\beta$ pairs are unstable (upper illustration). Therefore, the shear displacement occurs so that the diaphite structure are stabilized (lower illustration).}
  \label{fig:diaphite}
\end{center}
\end{figure}

Recently, using scanning tunneling microscopy (STM), Kanasaki {\it et al.} have discovered that the irradiation of visible photons onto the graphite induces novel nanoscale $sp^2\rightarrow sp^3$ collective conversion\cite{Kanasaki, Nature}. The STM image shows the nanoscale domain in which 4 carbons in each six-membered ring rise up from the layer and residual 2 sink down and close to the second layer, as shown in Fig.\ref{fig:diaphite}(a). The sinking down carbons form interlayer $\sigma$ bonds between the first and the second layers like Fig.\ref{fig:diaphite}(b). Due to the $AB$ stacking of the graphite, there are two types of interlayer bonded carbon pairs, namely, $\alpha$ and $\beta$ pairs. In order to stabilize these interlayer bonds, shear displacement occurs between the two layers. We briefly recapitulate the essential aspects of this experiment. The exciting light with an energy of 1.57 eV should be polarized perpendicular to the graphite layers and should be a femtosecond (fs) pulse. A polarization parallel to the layers or a picosecond (ps) pulse gives almost no contribution. Thus, this nonequilibrium phenomena requires an interlayer charge-transfer excitation and a transient generation of an excited electronic wave packet. Moreover, the process is nonlinear but at a lower level than a ten-photon process. The resultant domain is stable for several days at room temperature.

This phenomenon is closely related with graphite-diamond conversion, which has a long history of considerable experimental and theoretical researches\cite{Tateyama,Brenner,Fahy1,Fahy2,Bundy,Irifune,Banhart,Nakayama}. The diamond is about 0.02 eV/carbon higher than the graphite\cite{Brenner}, and the adiabatic energy barrier between them is about 0.3-0.4 eV/carbon\cite{Tateyama,Fahy1,Fahy2}. Therefore, to achieve uniform transformation from the graphite to the diamond, a macroscopic order of energy is required. For this reason, the conventional graphite-diamond conversion is well known to occur only under high temperature and pressure (3000 $^\circ$C, 15 GP)\cite{Bundy,Irifune}, or by the irradiation of strong X rays\cite{Banhart,Nakayama}.

Thus, the conventional conversion is quite different from the present photoinduced phenomenon. In the latter case, only a microscopic energy of several photons, enough to nucleate a minimum kernel of the new phase, is given to a local and limited area of the crystal. This minimum kernel will proliferate stepwise, according to the hidden stability intrinsic to the graphite. The photoinduced nanoscale domain shown in Fig. \ref{fig:diaphite} is not the conventional diamond, but an intermediate state between the graphite and the diamond, which is called ``diaphite.\cite{Kanasaki, Nature}''

From the above STM experiment, we can think of the following scenario for the present process\cite{Radosinski1}. When a fs-laser pulse is irradiated onto graphite layers perpendicularly, an electron-hole pair spanning two layers is generated. This electron-hole pair mainly dissipates into the  semimetallic continuum of the graphite as plus and minus carriers due to the good conductivity of graphite. However, by a small but finite probability, this electron-hole pair is expected to be bounded with each other through the interlayer Coulomb attraction. This exciton-like state self-localizes at a certain point of the layers by contracting the interlayer distance only around it. As the local contraction of the interlayer distance proceeds, an interlayer $\sigma$ bond is formed. Through further pulse excitations, an increasing amount of interlayer $\sigma$ bonds are formed stepwise, and then the diaphite structure is expected to appear macroscopically.

We have already clarified the adiabatic property of the phase transition by calculating the adiabatic path from the graphite to the small diaphite domain, by means of the LDA\cite{Ohnishi1} and by the semiempirical Brenner's theory\cite{Ohnishi2,Radosinski2} which can describe various carbon cluster systems. Next, we have studied the early stage dynamics of the interlayer $\sigma$ bond formation along with the above scenario\cite{Nishioka1,Radosinski2}. As a result, we found that the self-localization of an electron-hole pair spanning two layers occurs within a few femtoseconds at the probability of about 2\% when the electron-hole pair is excited as a transient state with the energy of 3.3 $\pm$ 1.8 eV. After the self-localization, by using classical molecular dynamics (MD) with the Brenner's potential, we obtained that the subsequent bond formation is achieved within about 0.5 ps when the excitation energy is more than 4.5 eV, corresponding to about three visible photons with the energy of 1.57 eV.

\begin{figure}
\begin{center}
  \includegraphics[width=1.0\linewidth]{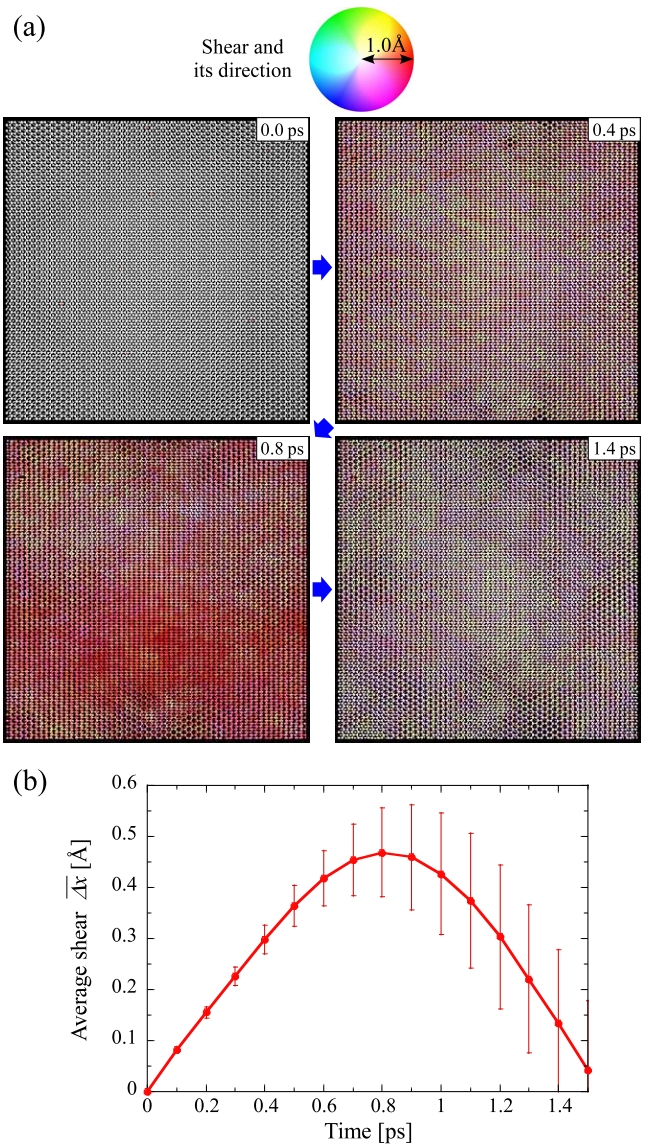}
  \caption{(a) The snapshots of the MD calculation in the case of 6\% excitation at t=0, 0.4, 0.8 and 1.4 ps. These are the top-down view of graphite layers. The system consists of two layers with $AB$ stacking and the initial interlayer distance 3.35 \AA. Each layer includes 6240 carbons in about 128 \AA\, $\times$ 128 \AA, wherein a periodic boundary condition is imposed. The shear between the upper layer and the lower one and its direction are represented by the circular color palette at the top. (b) The average shear displacement as a function of time. It is averaged over all carbon sites and over 50 calculations.}
  \label{fig:shear}
\end{center}
\end{figure}

Moreover, we investigated the effects of random multisite excitation by using the MD\cite{Nishioka2}. In this calculation, we select the excited sites randomly according to excitation density, and give random energy to each excited site according to the Gauss distribution with the mean energy of 3.3 eV and the width of 1.8 eV. Consequently, due to the cooperative phenomena between excited sites, the number of formed interlayer bonds increases nonlinearly with respect to the number of excited sites. The nonlinearity shows 1.7 power of the number of initially excited sites, indicating three or four photons are effectively involved in this interlayer bond formation. This theoretical result is also confirmed qualitatively by the experiment\cite{Kanasaki}.

However, the structure of the obtained domain with interlayer $\sigma$ bonds is quite different from the diaphite. In our calculation, all the interlayer bonds have been formed only in $\alpha$ pairs (we call them $\alpha$ bonds), not in $\beta$ pairs ($\beta$ bond) at all. As shown in Fig. \ref{fig:diaphite}(b), the diaphite includes $\alpha$ and $\beta$ bonds alternatively. The $\beta$ bonds are quite unstable due to the $AB$ stacking, so that they have not been formed in our calculation. As mentioned above, the $\beta$ bonds are stabilized by the shear displacement between layers like the lower structure of Fig. \ref{fig:diaphite}(b). The existence of the shear is particularly important for the formation of the diaphite domain, as already discussed in our previous works\cite{Kanasaki,Radosinski1,Ohnishi1,Ohnishi2,Radosinski2,Nishioka1,Nishioka2}. It has been experimentally confirmed before\cite{Kanasaki,Mishina} that the shear displacement is generated under fs-laser irradiation. However, it has not yet been clearly known how the shear occurs and $\beta$ bonds are successfully formed.

We discussed in our previous paper that when the graphite is excited only at $t=0$ the shear once appears due to the excitation of $\beta$ pairs but disappears with time, indicating that no $\beta$ bond is formed\cite{Nishioka2}. Fig. \ref{fig:shear}(a) shows the snapshots of the MD calculation in the random multisite excitation, representing the shear displacement of the upper layer relative to the lower one together with its direction by a circular color palette at the top. The excitation density is 6 \%, corresponding to 374 carbon pairs' excitation. In this calculation, assuming only the $x$-polarized light irradiation in $xy$ plane, all the excited $\beta$ pairs have the shift only in the $x$ direction.

As seen from these snapshots, the shear in $+x$ direction occurs in all the area once but afterward it disappears. Fig. \ref{fig:shear}(b) is the average shear displacement $\overline{\Delta x}\equiv N_{cal}^{-1}\sum_{i=1}^{N_{cal}}\left[N_{site}^{-1}\sum_{j=1}^{N_{site}}\Delta x_j\right]$, as a function of time. Here, $N_{cal}$ is the number of calculations (=50), $N_{site}$ the number of carbon sites in each layer, and $\Delta x_j\equiv x_{{\rm u},j}-x_{{\rm l},j}$, where $x_{{\rm u(l)},j}$ is the $x$ coordinate of the $j$ th site in the upper (lower) layer. In the 6\% excitation case, the average shear grows maximum at about 0.8 ps. Thus, the shear is proved to be only a transient and unfrozen one, unless the $\beta$ bonds are formed. However, since the shear makes $\beta$ bonds stable, they are expected to be formed more efficiently if the graphite were excited again before this transient shear disappears. In order to confirm it, we here investigate the effect of two-pulse excitation, using the MD.

\begin{figure}
\begin{center}
  \includegraphics[width=1.0\linewidth]{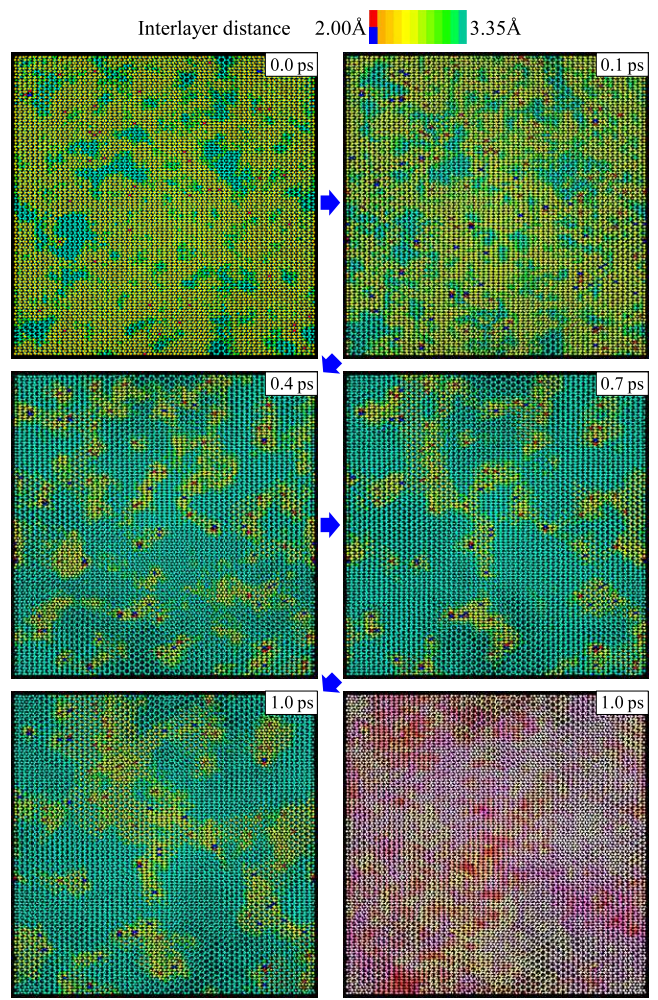}
  \caption{The snapshots of the MD calculation at t=0.4, 0.5, 0.8, 1.1 and 1.4 ps in the case where the additive excitation with 6\% density is given to the system at $t$=0.4 ps. The distance between the two layers is represented by the color palette at the top. The $\alpha$ and $\beta$ bonds are colored by red and blue, respectively. Only the snapshot at the bottom right shows the shear at $t$=1.4 ps.}
  \label{fig:addex_snapshot}
\end{center}
\end{figure}

The calculation procedure and the system are same as in the random multisite excitation case. After a certain period of time $\Delta t$ passes from the first excitation, we give additive excitation to the system. The excitation density of the second excitation is same as that of the first excitation, where we consider only 6\% excitation which is enough to induce cooperative phenomena in the interlayer bond formation\cite{Nishioka2}. We count out the number of $\alpha$ and $\beta$ bonds at 1.0 ps after the second excitation. By performing the calculations for various values of the time interval $\Delta t$, we investigate the $\Delta t$ dependence on the number of these two types of bonds.

As a example, we show the snapshots of the MD calculation for the case of $\Delta t=0.4$ ps in Fig. \ref{fig:addex_snapshot}. These snapshots represent the interlayer distance by a color palette at the top, where the $\alpha$ and $\beta$ bonds are colored by red and blue, respectively. Only the snapshot at the bottom right represents the shear displacement at 1.4 ps. As seen from this figure, although there is almost no $\beta$ bond until 0.4 ps, we can see a lot of $\beta$ bonds after the second excitation at 0.4 ps. These $\beta$ bonds tend to be formed near $\alpha$ bonds, since the interlayer distance is contracting there. Even at 1.4 ps, some of them remain together with $\alpha$ bonds, resulting in the small domains in which the interlayer distance is contracted. Moreover, as shown in the snapshot of the shear at the bottom right, the color of the domain including $\beta$ bonds remains red. This means that the shear does not disappear with time, and remains around the $\beta$ bonds. In other words, the transient shear is frozen by the second excitation. Thus, we confirmed that $\beta$ bonds can be formed due to the two-pulse excitation.

\begin{figure}
\begin{center}
  \includegraphics[width=0.9\linewidth]{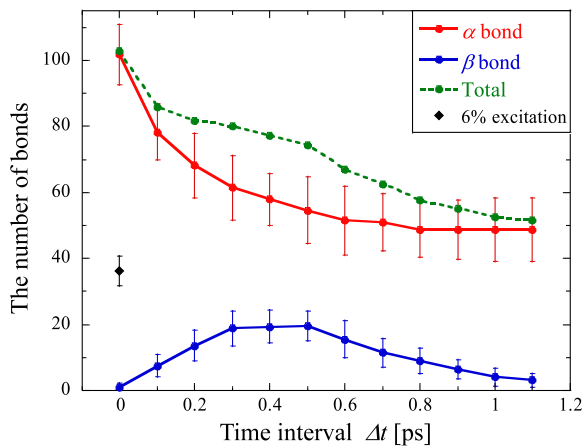}
  \caption{The number of interlayer bonds counted at 1.0 ps after the second excitation, as a function of the time interval $\Delta t$ between the first excitation and the second one. The red and blue solid lines indicate the number of $\alpha$ and $\beta$ bonds, respectively, and the green dashed one is the sum of them. The values at $\Delta t$=0 mean no additive excitation, corresponding to the 12\% excitation only at once. The black rhombus at $\Delta t$=0 corresponds to the 6\% excitation only at once.}
  \label{fig:nbonds}
\end{center}
\end{figure}

Fig. \ref{fig:nbonds} shows the $\Delta t$ dependence on the number of interlayer bonds, where the red and blue solid lines indicate the number of $\alpha$ and $\beta$ bonds, respectively, and the green dashed one is the sum of them. The values at $\Delta t$=0 mean no additive excitation, corresponding to the 12\% excitation only at once. For reference, the black rhombus at $\Delta t$=0 indicates the number of $\alpha$ bonds for the 6\% excitation at once. As seen from this figure, the number of $\alpha$ bonds monotonically decreases and settles down to a constant at more than $\Delta t$=0.9 ps, while $\beta$ bonds increase until $\Delta t$=0.5 ps and afterward decreases gradually. In this calculation, the $\beta$ bonds are formed most efficiently around $\Delta t$=0.3 - 0.5 ps. In this time region, the average shear is most rapidly increasing as seen from Fig. \ref{fig:shear}(b). Although the number of $\beta$ bonds are still quite less than that of $\alpha$ bonds even in this region, it is essential that $\beta$ bonds can be formed by the additive excitation. Because, if the $\beta$ bonds are once formed, the shear also remains stable around them. Therefore, through further excitation, new $\beta$ bonds will be formed more easily around the existing shear. Thus, $\beta$ bonds will proliferate comparable to $\alpha$ bonds will proliferate. As the number of these two types of bonds increases enough, the diaphite will be achieved. 

In conclusion, we have theoretically studied the nonequilibrium formation process of $sp^3$-bonded carbon nanodomains with shear displacement between graphite layers by means of the MD calculation. By the fs-laser excitation of graphite, the interlayer contraction and in-plane shear displacement are generated only as coherent and transient lattice motions. These lattice motions are getting frozen as interlayer $sp^3$ bonds increase stepwise by the additive pulse excitations. We have thus proposed the two-pulse excitation to achieve the efficient transition from the graphite to the diaphite.

This work is supported by the Ministry of Education, Culture, Sports, Science and Technology of Japan, the peta-computing project, and Grant-in-Aid for Scientific Research (S), Contract No. 19001002, 2007.

\end{document}